\documentclass[prl,showpacs,superscriptaddress,floatfix,twocolumn]{revtex4}
\usepackage{amsfonts}
\usepackage{stmaryrd}
\usepackage{bbm}
\usepackage{mathrsfs}
\usepackage{tipa}
\usepackage{amssymb}
\usepackage{txfonts}
\usepackage{graphicx}
\usepackage{dcolumn}
\usepackage{epstopdf}
\usepackage{array}

\newcommand{\PreserveBackslash}[1]{\let\temp=\\#1\let\\=\temp}
\newcolumntype{C}[1]{>{\PreserveBackslash\centering}p{#1}}
\newcolumntype{R}[1]{>{\PreserveBackslash\raggedleft}p{#1}}
\newcolumntype{L}[1]{>{\PreserveBackslash\raggedright}p{#1}}

\begin{document}

\newcommand*{\cm}{cm$^{-1}$\,}

\title{Observation of Open-Orbit Fermi Surface Topology in Extremely Large Magnetoresistance Semimetal MoAs$_2$}

\author{R. Lou}
\affiliation{Department of Physics and Beijing Key Laboratory of Opto-electronic Functional Materials $\textsl{\&}$ Micro-nano Devices, Renmin University of China, Beijing 100872, China}

\author{Y. F. Xu}
\affiliation{Beijing National Laboratory for Condensed Matter Physics, and Institute of Physics, Chinese Academy of Sciences, Beijing 100190, China}
\affiliation{School of Physical Sciences, University of Chinese Academy of Sciences, Beijing 100190, China}

\author{L.-X. Zhao}
\affiliation{Beijing National Laboratory for Condensed Matter Physics, and Institute of Physics, Chinese Academy of Sciences, Beijing 100190, China}
\affiliation{School of Physical Sciences, University of Chinese Academy of Sciences, Beijing 100190, China}

\author{Z.-Q. Han}
\affiliation{Department of Physics and Beijing Key Laboratory of Opto-electronic Functional Materials $\textsl{\&}$ Micro-nano Devices, Renmin University of China, Beijing 100872, China}

\author{P.-J. Guo}
\affiliation{Department of Physics and Beijing Key Laboratory of Opto-electronic Functional Materials $\textsl{\&}$ Micro-nano Devices, Renmin University of China, Beijing 100872, China}

\author{M. Li}
\affiliation{Department of Physics and Beijing Key Laboratory of Opto-electronic Functional Materials $\textsl{\&}$ Micro-nano Devices, Renmin University of China, Beijing 100872, China}
\affiliation{Shanghai Synchrotron Radiation Facility, Shanghai Institute of Applied Physics, Chinese Academy of Sciences, Shanghai 201204, China}

\author{J.-C. Wang}
\affiliation{Department of Physics and Beijing Key Laboratory of Opto-electronic Functional Materials $\textsl{\&}$ Micro-nano Devices, Renmin University of China, Beijing 100872, China}

\author{B.-B. Fu}
\affiliation{Beijing National Laboratory for Condensed Matter Physics, and Institute of Physics, Chinese Academy of Sciences, Beijing 100190, China}
\affiliation{School of Physical Sciences, University of Chinese Academy of Sciences, Beijing 100190, China}

\author{Z.-H. Liu}
\affiliation{State Key Laboratory of Functional Materials for Informatic, SIMIT, Chinese Academy of Sciences, Shanghai 200050, China}

\author{Y.-B. Huang}
\affiliation{Shanghai Synchrotron Radiation Facility, Shanghai Institute of Applied Physics, Chinese Academy of Sciences, Shanghai 201204, China}

\author{P. Richard}
\affiliation{Beijing National Laboratory for Condensed Matter Physics, and Institute of Physics, Chinese Academy of Sciences, Beijing 100190, China}
\affiliation{School of Physical Sciences, University of Chinese Academy of Sciences, Beijing 100190, China}
\affiliation{Collaborative Innovation Center of Quantum Matter, Beijing, China}

\author{T. Qian}
\affiliation{Beijing National Laboratory for Condensed Matter Physics, and Institute of Physics, Chinese Academy of Sciences, Beijing 100190, China}
\affiliation{Collaborative Innovation Center of Quantum Matter, Beijing, China}

\author{K. Liu}
\affiliation{Department of Physics and Beijing Key Laboratory of Opto-electronic Functional Materials $\textsl{\&}$ Micro-nano Devices, Renmin University of China, Beijing 100872, China}

\author{G.-F. Chen}
\affiliation{Beijing National Laboratory for Condensed Matter Physics, and Institute of Physics, Chinese Academy of Sciences, Beijing 100190, China}
\affiliation{School of Physical Sciences, University of Chinese Academy of Sciences, Beijing 100190, China}
\affiliation{Collaborative Innovation Center of Quantum Matter, Beijing, China}

\author{H. M. Weng}
\affiliation{Beijing National Laboratory for Condensed Matter Physics, and Institute of Physics, Chinese Academy of Sciences, Beijing 100190, China}
\affiliation{Collaborative Innovation Center of Quantum Matter, Beijing, China}

\author{H. Ding}
\affiliation{Beijing National Laboratory for Condensed Matter Physics, and Institute of Physics, Chinese Academy of Sciences, Beijing 100190, China}
\affiliation{School of Physical Sciences, University of Chinese Academy of Sciences, Beijing 100190, China}
\affiliation{Collaborative Innovation Center of Quantum Matter, Beijing, China}

\author{S.-C. Wang}
\email{scw@ruc.edu.cn}
\affiliation{Department of Physics and Beijing Key Laboratory of Opto-electronic Functional Materials $\textsl{\&}$ Micro-nano Devices, Renmin University of China, Beijing 100872, China}

\begin{abstract}
  While recent advances in band theory and sample growth have expanded the series of extremely large magnetoresistance (XMR) semimetals in transition
  metal dipnictides $TmPn_2$ ($Tm$ = Ta, Nb; $Pn$ = P, As, Sb), the experimental study on their electronic structure and the origin of XMR is still
  absent. Here, using angle-resolved photoemission spectroscopy combined with first-principles calculations and magnetotransport measurements, we
  performed a comprehensive investigation on MoAs$_2$, which is isostructural to the $TmPn_2$ family and also exhibits quadratic XMR. We resolve
  a clear band structure well agreeing with the predictions. Intriguingly, the unambiguously observed Fermi surfaces (FSs) are dominated by an
  open-orbit topology extending along both the [100] and [001] directions in the three-dimensional Brillouin zone. We further reveal the trivial
  topological nature of MoAs$_2$ by bulk parity analysis. Based on these results, we examine the proposed XMR mechanisms in other semimetals, and
  conclusively ascribe the origin of quadratic XMR in MoAs$_2$ to the carriers motion on the FSs with dominant open-orbit topology, innovating in
  the understanding of quadratic XMR in semimetals.
\end{abstract}


\maketitle

The emergence of novel states in condensed matter is not only classified by the typical spontaneous symmetry breaking, but also by their topology, i.e.,
the topologically protected quantum states \cite{Hasan2010,Qi2011,Weng2014}. The discovery of such symmetry protected states of matter in two-dimensional
(2D) \cite{Bernevig2006,Konig2007,Knez2011} and three-dimensional (3D) topological insulators \cite{Chen2009}, node-line semimetals \cite{Burkov2011,Yu2015},
topological crystalline insulators \cite{Fu2011,Heish2012NC}, and Dirac and Weyl semimetals \cite{Wang2012,Liu2014,Weng2015Weyl,Lv2015,Huang2015,Xu2015Weyl},
has attracted tremendous interests in condensed matter physics and material science. The magnetotransport behavior of these states is often unusual, such as
linear transverse magnetoresistance (MR) and negative longitudinal MR in Dirac and Weyl semimetals \cite{FengJ2015,Liang2015,HuangX2015,Shekhar2015,Xiong2015,
ZhangC2016,LiC2015}, and more generally, extremely large transverse MR (XMR) in nonmagnetic semimetals \cite{Schubnikow1930,Alers1951,Kasuya1993,Yang1999,
XuR1997,Mun2012}.

Recently, the discovery of XMR in a class of transition metal dipnictides $TmPn_2$ ($Tm$ = Ta, Nb; $Pn$ = P, As, Sb) \cite{WangK2014,ShenB2016,WuD2016,
XuC2016,WangY2016,WangZ2016} has sparked immense interests for understanding the underlying mechanism of quadratic XMR and exploring novel quantum states
arising from nontrivial topology. Another two series of semimetals possessing quadratic XMR behavior and rich topological characteristics are the ZrSiS
family \cite{Singha2017,Ali2016,WangXF2016} and Ln$X$ (Ln = La, Y, Nd, or Ce; $X$ = Sb/Bi) series \cite{Tafti2015,Sun2016,Kumar2016,Yu2016,Pavlosiuk2016,
Wakeham2016,Alidoust2016}, whose electronic structures have been considerably studied both in theory and experiment \cite{XuQ2015,Lou2016ZST,Schoop2016,
Zeng2015,Lou2016LaSb,Lou2017LaBi,Guo2016LnX}. While the band structures of the $TmPn_2$ series have been theoretically characterized in several work \cite{
XuC2016,ShenB2016,WuD2016,Luo2016}, experimental observations have not yet been reported. It is widely believed that the large positive MR in semimetals
is intimately related to their underlying electronic structures. Therefore, a systematic and unambiguous experimental study on the electronic structure
of the $TmPn_2$ family is urgently demanded. Eventually, we suggest the open-orbit Fermi surface (FS) topology as another candidate mechanism to explain
the XMR, in addition to the earlier proposed origins like nontrivial band topology \cite{Tafti2015}, forbidden backscattering at zero field \cite{JiangJ2015},
and electron-hole compensation \cite{Ali2014}.

\begin{figure}[htb]
  \begin{center}
  \includegraphics[trim = 4.5mm 0mm 0mm 0mm, clip=true, width=1.05\columnwidth]{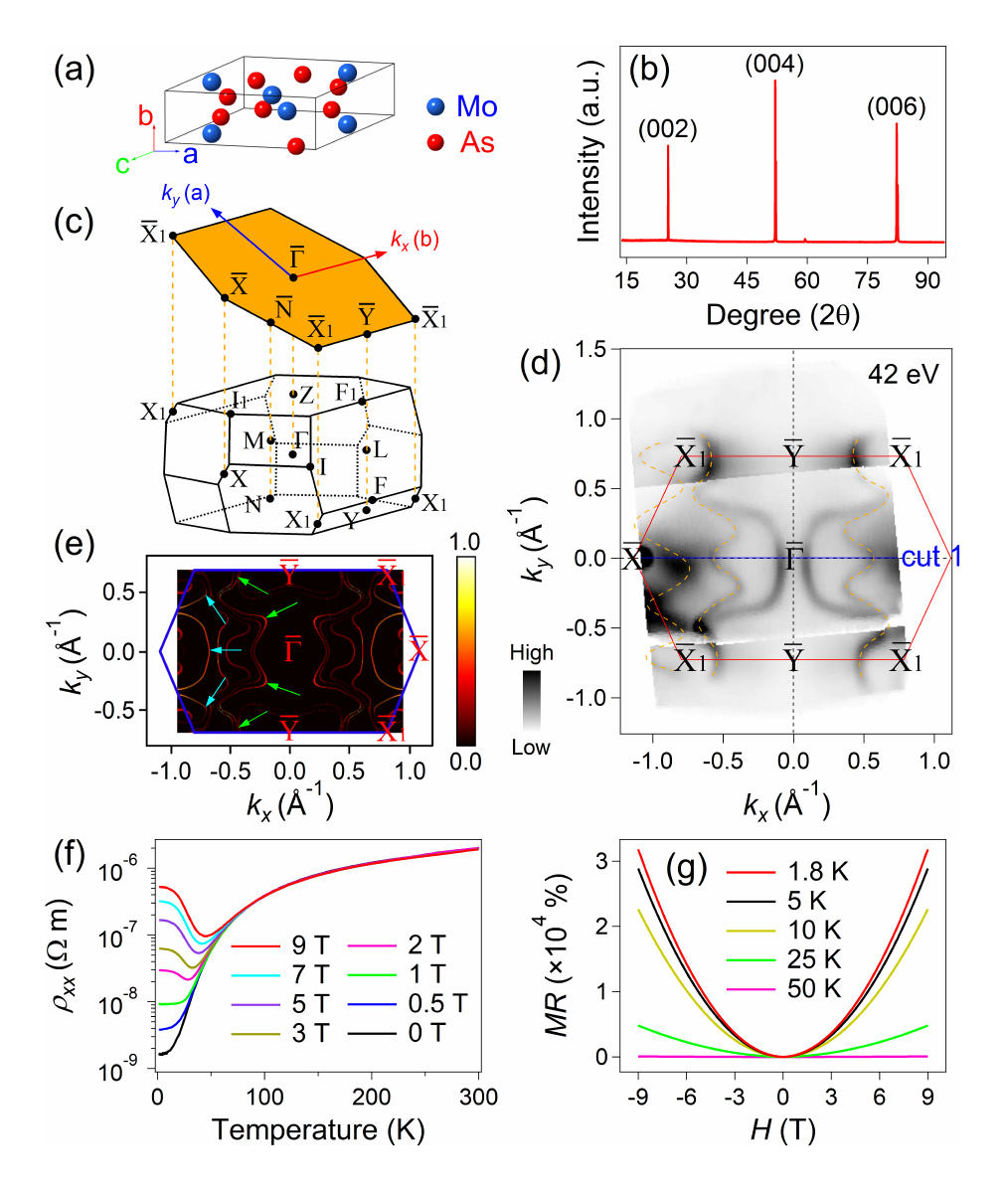}
  \end{center}
  \caption{
  (a) Schematic crystal structure of MoAs$_2$.
  (b) XRD pattern on the (001) surface.
  (c) Schematic primitive BZ and 2D projected BZ of the (001) surface.
  (d) Constant energy ARPES image obtained by $h\nu$ = 42 eV at $E_F$ with the mapping center around $\bar{\Gamma}$ point. Cut 1 indicates the momentum
      location of the measured bands in Figs. 2(a) and 2(b). Red lines represent the (001) surface BZ. Orange dashed curves are guides to the eye for the
      ``double ripple"-shaped FSs.
  (e) Calculated FSs for a 10-unit-cell-thick (001) slab with the MoAs-terminated layer. The inner and outer ``ripple"-shaped FSs are indicated by green
      and cyan arrows, respectively. Blue lines represent the (001) surface BZ. The intensity of the color scales the spectral weight projected to the
      topmost unit cell.
  (f) Temperature dependence of the resistivity in different magnetic fields. The magnetic field is parallel to the [001] direction and the electric current
      is parallel to the [010] direction ($b$-axis).
  (g) $MR$ (\%) = [$R$($H$) - $R$(0)]/$R$(0) $\times$ 100 \% plotted as a function of magnetic field at temperatures from 1.8 to 50 K.}
\end{figure}

In this Letter, we employ systematic angle-resolved photoemission spectroscopy (ARPES), first-principles calculations, and magnetotransport
measurements on MoAs$_2$, which is isostructural to the $TmPn_2$ family and exhibits quadratic MR exceeding 3.2 $\times$ 10$^4$\% at 1.8 K
under 9-T magnetic field. The FS topology is clearly resolved to display two ``double ripple"-shaped FSs extending along the [100] direction
($a$-axis), which do not close along the [001] direction (normal to the $ab$-plane) either, and one pocket at $\bar{X}$. Besides, we identify
two ``handle"-like FSs around $\bar{\Gamma}$, which arise from a trivial massless surface state (SS) along $\bar{\Gamma}$-$\bar{X}$. Our detailed
electronic structure of MoAs$_2$ would facilitate a more comprehensive understanding of the quadratic XMR in the $TmPn_2$ family.

\begin{figure*}[htb]
  \begin{center}
  \includegraphics[width=1.8\columnwidth]{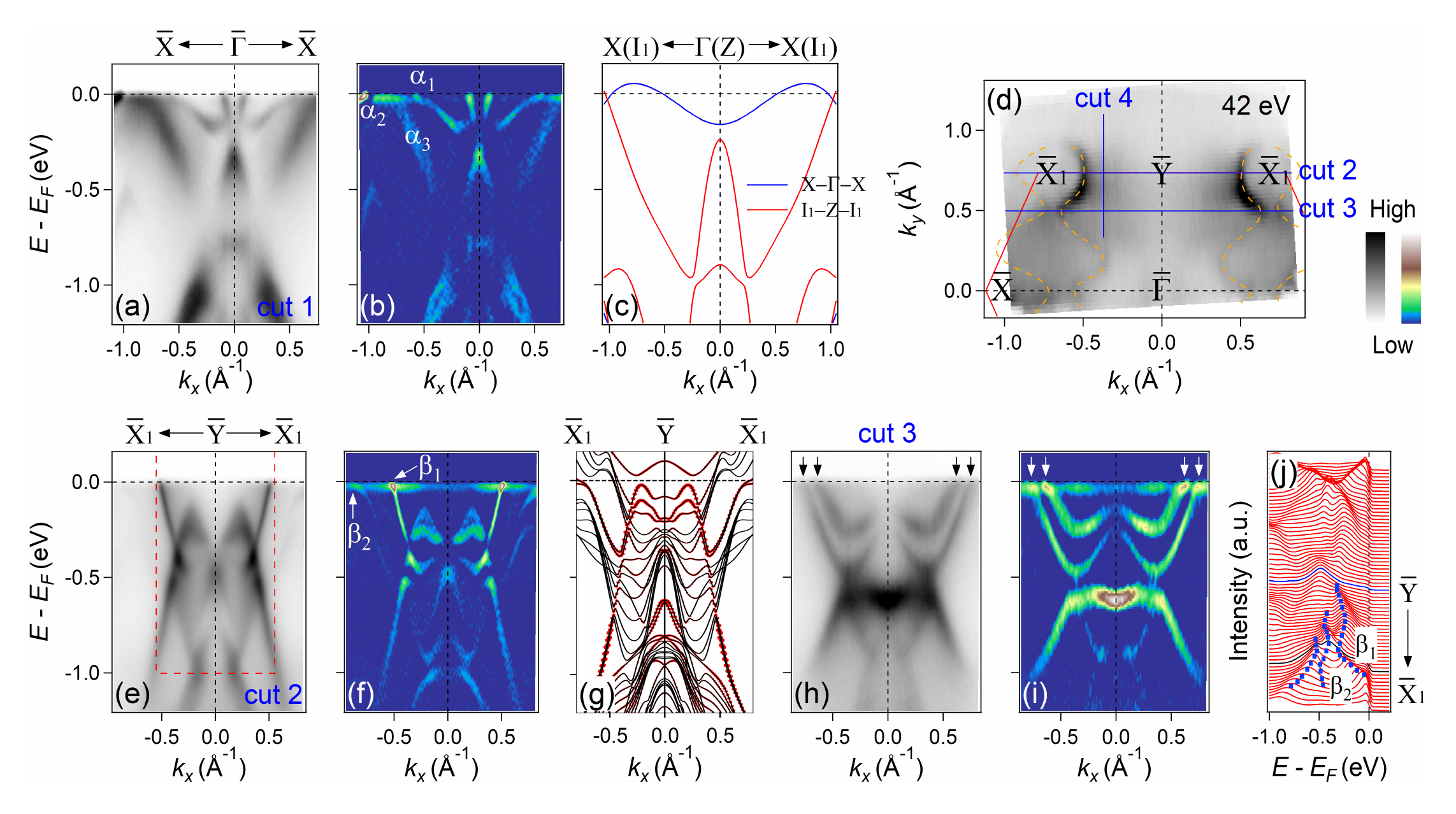}
  \end{center}
  \caption{
  (a),(b) Photoemission intensity plot and corresponding second derivative plot along $\bar{\Gamma}$-$\bar{X}$ [cut 1 in Fig. 1(d)], respectively.
  (c) Calculated band structures along $\Gamma$-$X$ and $Z$-$I_1$, considering that $X$ and $I_1$ points are approximately projected to one point
      on the (001) surface.
  (d) FS intensity plot obtained by integrating the spectral weight within $E_F$$\pm$10 meV recorded with $h\nu$ = 42 eV, the mapping center is near
      $\bar{Y}$ point. Cuts 2 and 3 indicate the momentum locations of the experimental band structures in (e)-(j). The electronic structures measured
      along cut 4 are presented in Fig. S4 (Sec. V of SM \cite{Supplementary}).
  (e),(f) Same as (a),(b) but along $\bar{Y}$-$\bar{X}_1$ [cut 2 in (d)]. Red dashed rectangle illuminates the area for EDC analysis in (j).
  (g) Calculated band structure along $\bar{X}_1$-$\bar{Y}$-$\bar{X}_1$ for a 10-unit-cell-thick (001) slab.
  (h),(i) Same as (a),(b) but along cut 3 in (d).
  (j) EDC plot of (e). The band gap is highlighted by black curve. Blue dots are extracted peak positions, serving as guides to the eye.}
\end{figure*}

High-quality single crystals of MoAs$_2$ with large residual resistance ratio [RRR = $R$(300 K)/$R$(1.8 K) = 1238] were grown via the chemical vapor
transport method. Detailed methods for ARPES and first-principles calculations can be found in Sec. I of Supplemental Material (SM) \cite{Supplementary}.
MoAs$_2$ crystallizes in a monoclinic structure with space group $C2/m$ (No. 12), as illustrated in Fig. 1(a) \cite{WangJ2016}. It is isostructural to
the well-known $TmPn_2$ family. Figure 1(b) shows the x-ray diffraction (XRD) pattern recorded on single crystals measured by ARPES, indicating our
measurements were performed on the (001) plane. The schematic 3D Brillouin zone (BZ) of the primitive cell and the corresponding 2D projected BZ of
the (001) surface are presented in Fig. 1(c). In Fig. 1(d), one can see the FSs measured with $h\nu$ = 42 eV exhibiting obvious modulations along $k_y$.
We observe two ``double ripple"-shaped FSs elongated along the [100] direction near the BZ boundary $\bar{X}$ and $\bar{X}_1$, one pocket at $\bar{X}$,
as well as two ``handle"-like FSs around $\bar{\Gamma}$. By further comparing with the calculated bulk band dispersions illustrated in Fig. 2, one can
identify that the ``handle"-like FSs are not of bulk origin, but rather derive from a non-topologically protected SS as discussed in the following. The
experimental FSs are in good agreement with the slab model calculations in Fig. 1(e) \cite{AddFSs}.

Figure 1(f) is the temperature evolution of the resistivity in different magnetic fields. MoAs$_2$ shows the metallic behavior from 1.8 to 300 K
without magnetic field. When the 9-T magnetic field is applied, one can see a prominent up-turn and a saturation at low temperatures. As depicted
in Fig. 1(g), MoAs$_2$ exhibits rather large MR exceeding 3.2 $\times$ 10$^4$\% at 1.8 K under 9-T magnetic field, and there is a drastic suppression
of MR with increasing temperature. One can obtain a nearly-quadratic exponent $m$ = 1.97 by fitting the $MR$-$H$ profile at 1.8 K with a power-law
function. These fingerprints are consistent with the common features of XMR semimetals \cite{Ali2014}.

In order to uncover the underlying nature of quadratic XMR and topological characteristics in MoAs$_2$, we investigate the near-$E_F$ band dispersions
along $\bar{\Gamma}$-$\bar{X}$ and $\bar{Y}$-$\bar{X}_1$. The band structure along $\bar{\Gamma}$-$\bar{X}$ is presented in Figs. 2(a)-2(c). As shown
in Figs. 2(a) and 2(b), we identify one Dirac cone-like band centered at $\bar{\Gamma}$ that will be discussed below. Three additional bands crossing
$E_F$ are clearly distinguished. The innermost electron band ($\alpha_1$) disperses slowly and crosses $E_F$ at $k_x$ $\sim$ -0.58 $\AA^{-1}$, forming
the inner ``ripple"-shaped FS, then it turns back as the outmost electron band ($\alpha_2$), which is in agreement with the bulk band calculations along
$\Gamma$-$X$ depicted in Fig. 2(c). As the absence of the middle one ($\alpha_3$), which forms the outer ``ripple"-shaped FS, in the calculations along
$\Gamma$-$X$, we suggest that $\alpha_3$ may not come from the $k_z$ = 0 plane. Due to the short escape length of the photoelectrons excited by the vacuum
ultraviolet light in our ARPES experiments, the $k_z$ broadening effect would be prominent \cite{Strocov2003}, demonstrated by previous ARPES studies \cite{
Kumigashira1997,Kumigashira1998,Lou2016LaSb,Lou2017LaBi,Niu2016,DTakane2016}, we further perform bulk band calculations along $Z$-$I_1$ (in the $k_z$ = $\pi$
plane) as shown in Fig. 2(c). The consistency between $\alpha_3$ and the calculations helps finding out the origin of this band \cite{Morekz}.

\begin{figure}[htb]
  \begin{center}
  \includegraphics[width=0.88\columnwidth]{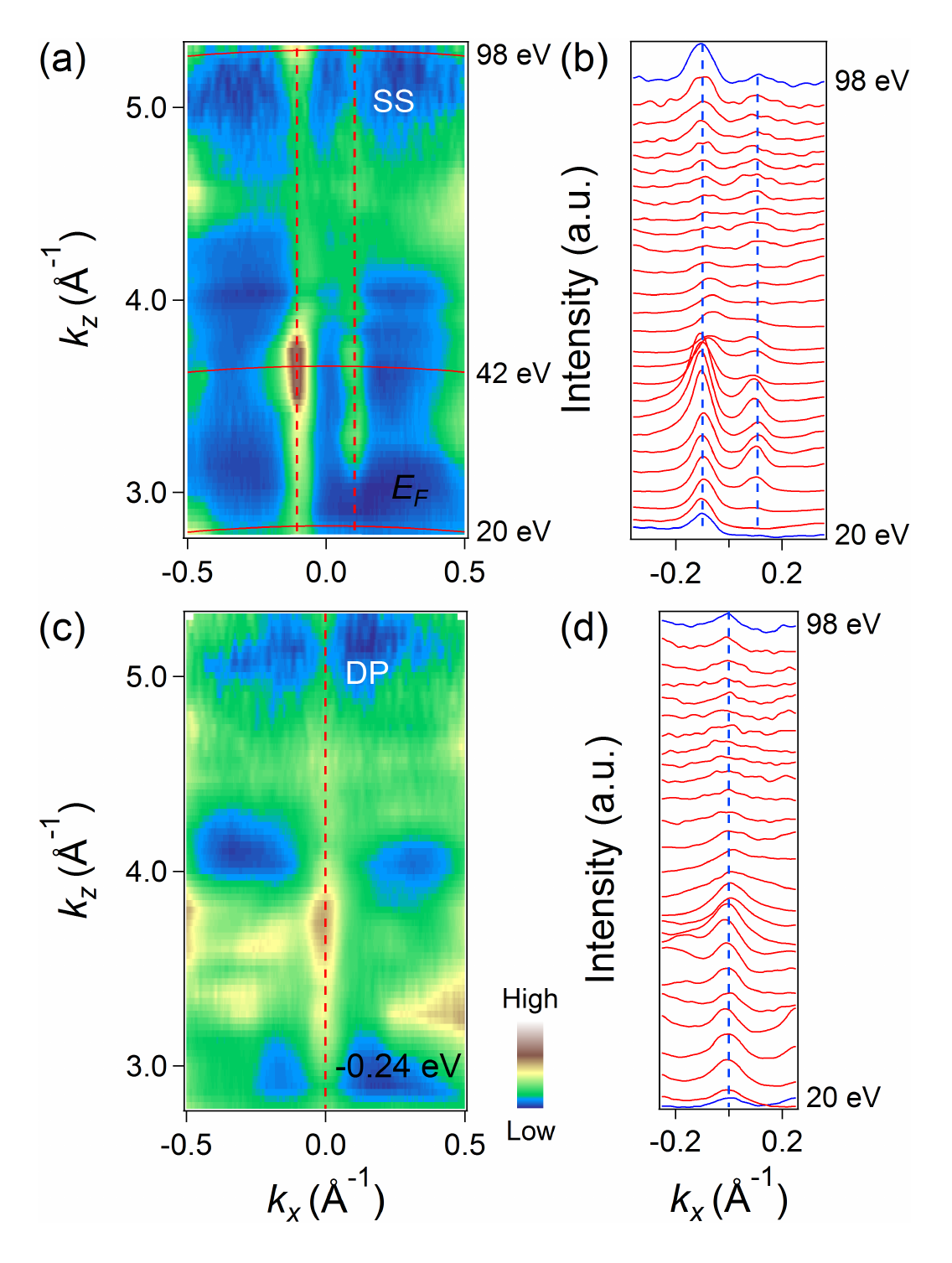}
  \end{center}
  \caption{
  (a) ARPES intensity map at $E_F$ in the $k_x$-$k_z$ plane at $k_y$ = 0 recorded with various photon energies.
  (b) MDCs of (a) around the BZ center.
  (c) Same as (a) but at $E$ = -0.24 eV.
  (d) MDCs of (c) around the BZ center.
  Red and blue dashed lines in (a),(b) and (c),(d) show the Fermi crossings and the DP of the SS around $\bar{\Gamma}$, where identifiable, respectively.}
\end{figure}

We investigate the electronic structure along $\bar{Y}$-$\bar{X}_1$ to examine the origin of the ``double ripple"-shaped FSs. As indicated in Fig. 2(d),
the ``double ripple"-shaped feature can be resolved more clearly from the FS mapping focusing around $\bar{Y}$ point, while the ``handle"-like FSs are
relatively weak in this geometry due to the matrix element effect. From the measured band dispersions along $\bar{Y}$-$\bar{X}_1$ illustrated in Figs.
2(e) and 2(f), one can observe two almost linearly dispersive bands crossing $E_F$ indicated by white arrows. These two bands are denoted as $\beta_1$
and $\beta_2$, which correspond to the inner and outer ``ripple"-shaped FSs, respectively. Detailed energy distribution curve (EDC) analysis on $\beta_1$
and $\beta_2$ are shown in Figs. 2(j) and S3 (Sec. IV of SM \cite{Supplementary}), respectively. The experimental band structures agree with the slab
calculations along $\bar{X}_1$-$\bar{Y}$-$\bar{X}_1$ shown in Fig. 2(g). By compared with the bulk calculations along $Y$-$X_1$ (near the $k_z$ = $\pi$
plane), displayed as Fig. S2(b) in Sec. III of SM \cite{Supplementary}, the fact that $\alpha_3$, which is captured by the calculations along $Z$-$I_1$
rather than $\Gamma$-$X$, derives from the projection of the $k_z$ $\approx$ $\pi$ plane due to the $k_z$ broadening effect \cite{Lou2016LaSb,Lou2017LaBi}
can be further supported. The consistency between measured electronic structures along $\bar{Y}$-$\bar{X}_1$ and theoretical calculations can be proved
by the presence of a band gap between $\beta_1$ and $\beta_2$ (see a quantitative analysis on this gap in Sec. V of SM \cite{Supplementary}). To further
demonstrate the open character of the in-plane FSs, we study the band structure along cut 3, i.e., $k_y$ $\sim$ 0.50 $\AA^{-1}$, the momentum where the
inner and outer ``ripple"-shaped FSs are closest to each other. In Figs. 2(h) and 2(i), as marked by vertical arrows, the two linearly dispersing bands
are clearly separated at $E_F$.

While most experimental band structures along $\bar{\Gamma}$-$\bar{X}$ are in agreement with the bulk band calculations, we observe an extra Dirac
cone-like band centered at $\bar{\Gamma}$. To illuminate the origin of this band, we investigate the band dispersions recorded with different photon
energies \cite{Damascelli2004,PRichard2015}. We show the constant energy plot in the $k_x$-$k_z$ plane at $k_y$ = 0 at $E_F$ and $E$ = -0.24 eV, which
is the binding energy of the Dirac point (DP) at $\bar{\Gamma}$, in Figs. 3(a) and 3(c), respectively. The corresponding momentum distribution curve
(MDC) plots are presented in Figs. 3(b) and 3(d), respectively. The Dirac-like band features, including the Fermi crossings and the DP, do not show
noticeable change with a varying photon energy over a wide range, confirming that it is a SS \cite{SS}.

\begin{figure}[htb]
  \begin{center}
  \includegraphics[width=0.97\columnwidth]{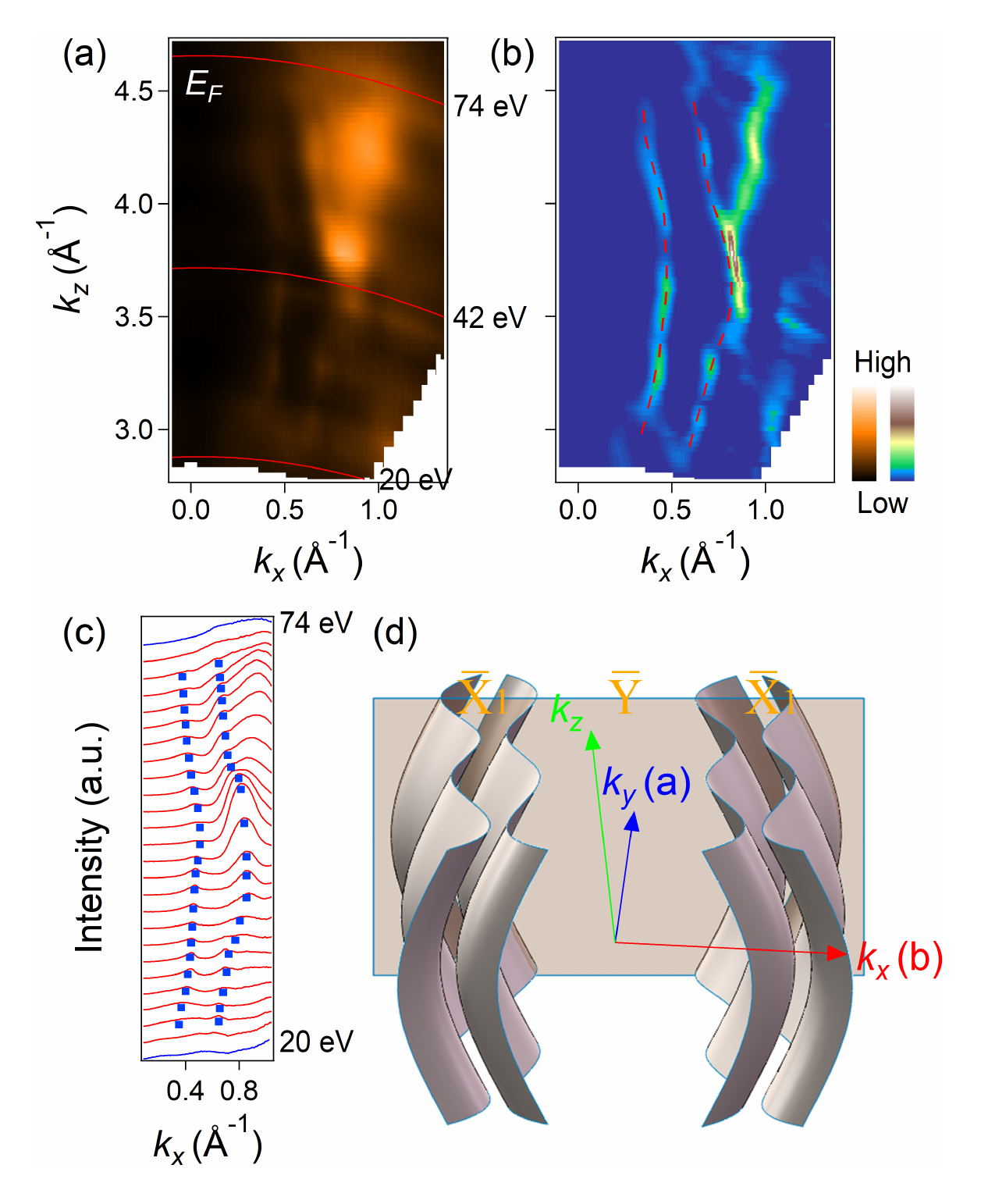}
  \end{center}
  \caption{
  (a) ARPES intensity map at $E_F$ in the $k_x$-$k_z$ plane at $k_y$ = $\pi$ recorded with various photon energies.
  (b) Second derivative plot of (a).
  (c) MDCs of (a). Red dashed curves in (b) and blue dots in (c), which are extracted peak positions, indicate the Fermi crossings of $k_z$ bandmap.
  (d) Schematic diagram of the open FSs extending in the 3D BZ. The translucent plane indicates the momentum location of the $k_x$-$k_z$ mapping in (a).}
\end{figure}

We further perform $k_z$-dependent measurements along $\bar{Y}$-$\bar{X}_1$ focusing on the ``double ripple"-shaped FSs. The photon energy variation
covers more than one BZ along the $k_z$ direction ($\sim$0.88 $\AA^{-1}$), which is sufficient to illustrate the periodicity along $k_z$. As shown in
Figs. 4(a)-4(c), the ``double ripple"-shaped open FSs do not close along the $k_z$ direction either (see Fig. S5 in Sec. VI of SM for more manifestations
of the open character \cite{Supplementary}). When the FS topology contains open orbits along certain directions, MR in materials is not saturated and
parabolically dependent on the magnetic field \cite{Abrikosov1988,Singleton2001}. To elucidate the possible electronic origin of quadratic XMR in MoAs$_2$, angular-dependent MR measurements were carried out (see details of the experimental setup and discussions on the anisotropy of MR in Sec. VIII of SM
\cite{Supplementary}).

So far, immense work has been generated on the origin of quadratic XMR behavior, and several mechanisms have been proposed, including nontrivial
band topology \cite{Tafti2015}, forbidden backscattering at zero field \cite{JiangJ2015}, and electron-hole compensation \cite{Ali2014}. First,
the field-induced resistivity upturn and plateau is suggested as the consequences of breaking time-reversal symmetry in topological semimetals
\cite{Tafti2015}, whereas the $Z_2$ classification of MoAs$_2$ is (0; 000) \cite{FuL2007} based on our first-principles calculations (detailed
analysis can be found in Sec. VII of SM \cite{Supplementary}), demonstrating that MoAs$_2$ is a topologically trivial material. This excludes
the possibility that the XMR in MoAs$_2$ is associated with nontrivial band topology. Second, the spin degeneracy is removed by spin-orbit coupling
due to the lack of inversion symmetry in WTe$_2$, leading to an exotic spin texture of the bands, which has been claimed to play an important role
in the XMR of WTe$_2$ \cite{JiangJ2015}. This explanation is not applicable to MoAs$_2$ either, because the inversion symmetry in MoAs$_2$ preserves
that its spins are doubly degenerate at zero field. Third, while electron-hole compensation can be broadly applied in many XMR semimetals with closed
FS trajectories based on the two-band model \cite{Lou2016LaSb,Sun2016,Ali2014,Pippard1989}, its availability in XMR materials with open FSs is still
unclear \cite{Ashcroft1976}. From another perspective, it should be noted that the MR tends to a constant, i.e., it saturates, in high fields for FSs
consisting of closed electron and hole pockets, unless they have exactly equal volume. This contrasts with the MR for open FSs, which increases as $H^2$
and exhibits non-saturation without the restriction of carrier compensation \cite{Ziman1972}. For the former, under the influence of a magnetic field,
carriers in clean materials with closed FSs will travel several orbits before getting scattered, resulting in velocity average to zero in the plane
perpendicular to the magnetic field. As for the carriers on open FSs, for which the product of the cyclotron frequency ($\omega_c$) and the relaxation
time ($\tau$) is no longer much larger than 1, thus leading to a finite in-plane velocity. These two conditions cause above two distinct field-dependence \cite{Abrikosov1988,Singleton2001}. Since the dominant open-orbit FS topology can, by itself, lead to the quadratic XMR in MoAs$_2$, while the dominant
closed FSs are necessary for the non-saturated MR induced by carrier compensation, thus, electron-hole compensation mechanism cannot be applied to the
XMR in MoAs$_2$.

Hence, it is reasonable to suggest that the quadratic XMR in MoAs$_2$ is attributed to the carriers motion on the FSs with dominant open-orbit
topology. Furthermore, despite what topology the FSs are, the ultra-high charge mobilities are the key condition effectively enhancing the MR
in semimetals \cite{Sun2016,Singleton2001}, this is also applicable in MoAs$_2$, of which $\mu_e$ and $\mu_h$ is 1.04 and 3.49 $\times$ 10$^4$
cm$^2$V$^{-1}$s$^{-1}$, respectively. They are determined by fitting the Hall conductivity at $T$ = 1.8 K using the two-carrier model \cite{
Takahashi2011,XiaB2013}, as illustrated in Fig. S1(b) (Sec. II of SM \cite{Supplementary}). The open FS character along certain directions
has also been observed in PdCoO$_2$ and ZrSiS exhibiting the XMR behavior \cite{Takatsu2013,Eyert2008,LvY2016}. However, for PdCoO$_2$, the
field-dependence of MR is not parabolic and the role of open FSs on the XMR has not yet been studied \cite{Takatsu2013}; for ZrSiS, the carrier
compensation is proposed to be dominant in the quadratic XMR when $H$ $\parallel$ [001], while the open-orbit FS topology serves to enhance the
MR ($\sim$50\%) when $H$ $\parallel$ [011], as the existence of open FSs when $H$ $\parallel$ [010] \cite{LvY2016}.

In conclusion, we present the comprehensive electronic structure of XMR semimetal MoAs$_2$. The observed FSs are dominated by open-orbit topology
extending along both the [100] and [001] directions. We demonstrate the trivial topological nature of MoAs$_2$ by bulk parity analysis. Our results
unambiguously suggest that the origin of quadratic XMR in MoAs$_2$ is attributed to the carriers motion on the FSs with dominant open-orbit topology,
serving as a novel mechanism for XMR in semimetals.

\begin{acknowledgments}
We would like to thank H. C. Lei, T.-L. Xia for fruitful discussions, and Y.-Y. Wang, S.-S. Sun for the help in transport measurements.
This work was supported by the Ministry of Science and Technology of China (Programs No. 2013CB921700, No. 2015CB921000, No. 2016YFA0300300,
No. 2016YFA0300600, No. 2016YFA0302400, and No. 2016YFA0401000), the National Natural Science Foundation of China (Grants No. 11774421, No.
11274381, No. 11274362, No. 11474340, No. 11234014, No. 11274367, No. 11474330, No. 11674371, and No. 11704394), and the Chinese Academy of
Sciences (CAS) (Project No. XDB07000000). R.L., K.L., and J.W. were supported by the Fundamental Research Funds for the Central Universities,
and the Research Funds of Renmin University of China (RUC) (Grants No. 17XNH055, No. 14XNLQ03, and No. 17XNLF06). Z.L. was supported by the
Shanghai Sailing Program (Grant No. 17YF1422900). Y.H. was supported by the CAS Pioneer Hundred Talents Program.

R. L., Y. F. X., L.-X. Z., and Z.-Q. H. contributed equally to this work.
\end{acknowledgments}


\begin{thebibliography}{1}
\bibitem{AddFSs}
Compared with the experimental FS topology, the additional FSs in slab calculations possibly come from the hybridization between the wave
functions of the topmost unit cell and the rest unit cells, because the coherence length of atomic wave functions can reach to multiple unit
cells. The projected spectral weight of these states is relatively weak and thus cannot be clearly identified in the experiments. The matrix
element effect may also contribute to this issue to some extent.
\bibitem{SS}
The SS is not topologically protected because of the trivial $Z_2$ index discussed below and may be merged into the bulk continuum spectra
by small surface perturbations. Thus, the SS is not well reproduced in the calculated FSs in Fig. 1(e).
\end{thebibliography}
\end{document}